\documentclass[twocolumn,aps,showpacs,preprintnumbers,superscriptaddress,amsmath,amssymb,prl]{revtex4-1}  
\usepackage{graphicx}  
\usepackage{dcolumn}   
\usepackage{bm}        
\usepackage{color}

\usepackage{times}
\usepackage{amsmath}
\usepackage{epsfig}
\usepackage{epstopdf}
\usepackage{subfigure}
\usepackage{multirow}
\usepackage{siunitx}
\usepackage{color}

\newcommand{\figref}[1]{{Fig.~\ref{#1}}}

\def\acsphoto{ACS \ Photon. }
\def\acsnano{ACS\ Nano }
\def\aplphylet{Appl.\ Phys.\ Lett.}
\def\aplopt {Appl. Opt.}

\def\intjmodphyb{Int.\ J.\ Mod.\ Phys.\ B}
\def\jap{J.\ Appl.\ Phys. }
\def\jetlet{ JETP\ Lett. }
\def\lasphorev{ Laser\ Photon.\ Rev.}
\def\nat{ Nature}
\def\natmat{ Nat.\ Mater.}
\def\natpho{ Nat.\ Photon.}
\def\natphy{ Nat.\ Phys.}
\def\natnano{ Nat.\ Nanotechnol.}
\def\nanolet{ Nano \ Lett.}
\def\nanotod{ Nano Today}
\def\newjphy{ New J.\ Phys.}
\def\optexp{Opt.\ Express }
\def\optlet{Opt.\ Lett. }

\def\prl{ Phys.\ Rev.\ Lett.}
\def\prb{ Phys.\ Rev.\ B}
\def\prx{ Phys.\ Rev.\ X}
\def\ptrsa{ Phil.\ Trans.\ R.\ Soc.\ A }
\def\revmodphy{Rev.\ Mod.\ Phys. }
\def\sci{Science }
\def\scirep{Sci.\ Rep.}

\begin{document}

\title{Tunable and Dual-broadband Giant Enhancement of SHG and THG in a Highly-engineered Graphene-Insulator-Graphene Metasurface}
\date{\today}

\author{Jian~Wei~You}
\author{Nicolae~C.~Panoiu}
\affiliation{Department of Electronic and Electrical Engineering, University College London,
Torrington Place, London WC1E 7JE, United Kingdom}

\begin{abstract}
We demonstrate a novel scheme to dramatically enhance both the second- and third-harmonic
generation in a graphene-insulator-graphene metasurface. The key underlying feature of our approach
is the existence of a double-resonance phenomenon, namely the metasurface is designed to possess
fundamental plasmon resonances at both the fundamental frequency and the higher harmonic. In
particular, this dual resonant field enhancement at the two optical frequencies, combined with a
favorable spatial overlap of the optical near-fields, lead to the increase of the generated higher
harmonic by several orders of magnitude. Remarkably, we demonstrate that by tuning the Fermi energy
of the graphene gratings the dual-resonance property can be locked-in over a broad spectral range
of $\sim$\SI{20}{\tera\hertz}, and equally important, the enhanced nonlinear frequency generation
process can be readily switched in the same device between the second and third harmonic. This new
type of graphene metasurface could open up new avenues towards the development of novel
ultra-compact and multi-frequency active photonic nanodevices.
\end{abstract}

\maketitle

The first successful isolation of graphene from graphite \cite{ngm04sci} via mechanical exfoliation
has opened up a rapidly growing field of research
\cite{bsl06sci,gn07natmat,bgb08nanolet,han08jap,cgp09rmp}, primarily due to the unique and
remarkable properties of this new two-dimensional material. In its early stages research on
graphene focused on its electronic and mechanical properties, but it was soon realized that key
optical properties, such as extreme optical near-field confinement induced by the excitation of
surface-plasmon polaritons (SPPs)
\cite{gpn12natpho,cbg12n,fra12nat,aba14acsphoto,kca11nanolet,cba12nat,bfp13intjmodphyb}, tunability
of the optical response via gate voltage and chemical doping
\cite{lyu11nat,cry11nanotod,bjs13nanolet}, and low losses at high carrier densities
\cite{dyml10NatNano,la14AcsNano}, could transform graphene into a promising and versatile material
platform for a broad array of optoelectronic applications. To this end, photonic devices based on
graphene, including diffractive elements, optical sensors, topological photonic devices,
photovoltaic and photoresistive devices
\cite{bsh10natpho,bl12acsnano,ybbp18Nanophoto,ylet19NatNano,ldj10sci,kjbm18NanoLett,xml09natnano,sll10acsnano,pls10optexp,hygk19natcomm,jlbp20JSTQE,jcsf17PRL,pyxa17NatComm},
have already been demonstrated.

In addition to advances in exploiting the linear physics of graphene, its nonlinear optical
properties could play an equally important role in key applications. Due to the centrosymmetric
nature of graphene, the leading non-vanishing nonlinear optical interactions in graphene are of
third-order type, such as third-harmonic generation (THG) and Kerr effect. In particular, it has
been demonstrated that the strength of third-order nonlinear optical interactions in graphene is
several orders of magnitude larger than in typical semiconductors
\cite{hhm10prl,hdp13prx,cvs14newjphy,csa16acsnano}. More importantly, these nonlinear optical
interactions can be further enhanced upon resonant excitation of SPPs in graphene structures, which
leads to a number of exciting applications
\cite{bsh10natpho,bl12acsnano,ybbp18Nanophoto,wp16prb,yywp17ptrsa,vcg13optlet,
nan13lpr,dyp15prb,dcm15scirep,sgisk13prb,Soa18NatNano,snsk15prb,ylp20SciAdv}, including frequency
mixing \cite{bsh10natpho}, photodetectors \cite{xml09natnano}, generation of spatial solitons
\cite{nan13lpr,snsk15prb}, physical systems with tunable Dirac points \cite{dyp15prb}, and Anderson
light localization at the nanoscale \cite{dcm15scirep}.

Although the second-harmonic generation (SHG) is generally forbidden in a free-standing graphene
sheet, it is nevertheless permitted in two main configurations. First, SHG does arise in graphene
nanostructures from \textit{nonlocal} effects \cite{csa16acsnano,smki14prb,msa15njp}, namely when
nonlinear sources of SHG are magnetic dipoles and electric quandrupoles. Second, by placing a
graphene sheet on a substrate, the inversion symmetry of the system is broken and SHG due to
\textit{local} nonlinear polarization (electric dipoles) can occur
\cite{gla11jlet,mik11prb,chc16natphy,dh09apl,dh10prb,ard14prb}. Under these conditions, the
effective second-order susceptibility of graphene can be several orders of magnitude larger than
that of semiconductors widely used in nonlinear optics, e.g. GaAs \cite{b03book}.

In this Letter, we introduce a novel graphene-insulator-graphene (GIG) optical structure designed
to achieve a tunable and dual-broadband enhancement of both SHG and THG, and equally important, the
enhanced nonlinear frequency generation process can be readily switched in the same device between
the second-harmonic (SH) and third-harmonic (TH). The giant enhancement of both nonlinear optical
interactions is realized by ensuring that the GIG structure possesses \textit{first-order plasmon
resonances} at both the fundamental frequency (FF) and higher-harmonics (HHs), namely SH and TH.
For the sake of completeness, for SHG, we consider both the cases of a nonlocal nonlinear
polarization, which corresponds to graphene structures in a stacked configuration embedded in a
background medium \cite{ylc12natnano,imsbk13prb,sssk14LPR}, and the case of a local nonlinear
polarization, when graphene is placed on a substrate.

\begin{figure}[t!]\centering
\includegraphics[width=\columnwidth]{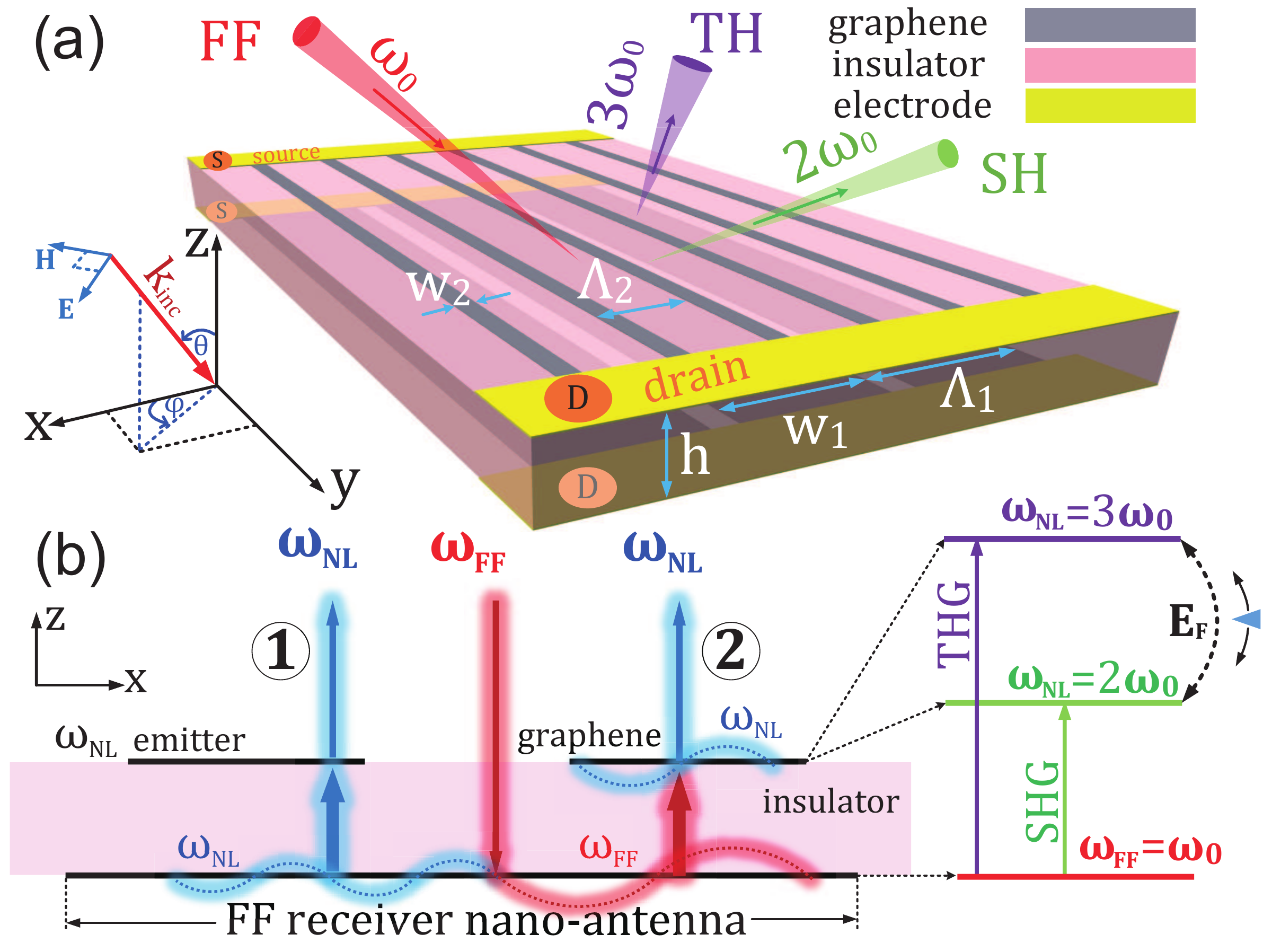}
\caption{(a) Schematic of a tunable GIG nanoresonator consisting of GNRs with different width
placed at the opposite facets of an insulator slab. (b) Illustration of physical mechanisms of
enhancement of SHG and THG in the GIG metasurface.}\label{fig:Geometry}
\end{figure}

The proposed periodic GIG structure is depicted in \figref{fig:Geometry}(a). Its unit cell consists
of two graphene nanoribbons (GNRs) placed at the opposite facets of an (insulator) dielectric
spacer. Electrodes are placed in contact with the GNRs, which allows one to tune their Fermi level.
A TM-polarized plane-wave with frequency $\omega_{0}$ is incident from above onto the GIG
structure. As SPPs of GNRs are geometry-dependent, their frequency can be set by properly choosing
the width of the ribbons. Using this feature, the widths of the GNRs are chosen in such a way that
the bottom and top GNRs have \textit{first-order} SPP resonances at both the FF,
$\omega_{FF}=\omega_{0}$, and HH ($\omega_{NL}=2\omega_{0}$ for SHG and $\omega_{NL}=3\omega_{0}$
for THG), respectively, as per \figref{fig:Geometry}(b). In addition, the nonlinear optical
response of the GIG structure can be further optimized by requiring that the bottom GNRs have
higher-order plasmons at the HH, too \cite{yywp17ptrsa}. Importantly, this nonlinear optical device
can be used to enhance both the SHG and THG by simply varying the Fermi level in the top GNRs, so
as the frequency of its first-order SPP is switched between $2\omega_{0}$ and $3\omega_{0}$.

There are two key mechanisms that contribute to the remarkably large enhancement of the nonlinear
optical response of the GIG structure, namely by several orders of magnitude as compared to that of
a graphene sheet. The first one, indicated by path \textcircled{1} in \figref{fig:Geometry}(b),
requires that the bottom GNR has a first-order plasmon at $\omega_{0}$ and a higher-order SPP at
$\omega_{NL}$ \cite{yywp17ptrsa}. Then, the field at $\omega_{0}$ incident onto the bottom GNRs
generates a strong field on these GNRs at $\omega_{0}$, via the resonant excitation of first-order
SPPs, and, subsequently, higher-order SPPs are resonantly generated by the nonlinear polarization
in these same bottom GNRs.

We now introduce a new, much more efficient mechanism contributing to the enhancement of the
nonlinear response of the GIG structure. It is schematically indicated by path \textcircled{2} in
\figref{fig:Geometry}(b) and relies on the fact that the top GNRs possess first-order SPPs at the
HH. This mechanism can be described as follows: the enhanced optical field due to the excitation of
first-order SPPs on the bottom GNRs induces on the top GNRs a strong nonlinear polarization at the
HH via near-field interaction. This, in turn, resonantly excites first-order SPPs on the top GNRs.
Additionally, first-order SPPs on the top GNRs (at HH) are also directly generated via optical
near-field coupling with higher-order SPPs of the bottom GNRs.

In the final stage of the nonlinear optical interaction between the incoming light and the GIG
structure, the higher-order SPPs on the bottom GNRs and the first-order SPPs on the top GNRs couple
to the radiative modes to generate a strong signal at the HH. In fact, this GIG system acts as a
nonlinear Yagi-Uda nanoantenna \cite{nn11natpho}: the bottom and top GNRs are the driver at
$\omega_{0}$ and director at $\omega_{NL}$, respectively.

To illustrate these ideas, we considered a metasurface with the periods of the bottom and top
graphene gratings of $\Lambda_1=\SI{200}{\nano\meter}$ and $\Lambda_2=\SI{100}{\nano\meter}$,
respectively. The widths $w_1$ and $w_2$ of the GNRs and the thickness, $h$, of the spacer are
designed so as to achieve a double-resonance effect. We assume that the spacer is made of
polyethylene, which has relative permittivity of $\epsilon_s=2.28$ and is practically lossless at
mid-infrared frequencies \cite{wam98ao}. The linear and nonlinear optical responses of this GIG
structure have been studied using an in-house developed code based on the GS-FDTD method; for
details on the numerical approach see the Supplemental Material (SM) \cite{SM17GSFDTD}.

In this method, the linear properties of graphene are modeled using a linear surface optical
conductivity \cite{Hans08jap},
\begin{equation}\label{eq:sigma}
\sigma_{s} = \frac{e^2 k_B T\tau}{\pi\hbar^2\overline{\omega}}\left[\frac{E_F}{k_B T} +2\ln
\left(e^{-\frac{E_F}{k_B T}} + 1\right) \right]
+\frac{ie^2}{4\pi\hbar}\ln\frac{\xi-i\overline{\omega}}{\xi+i\overline{\omega}}.
\end{equation}
Here, $E_F$, $T$, and $\tau$ are the Fermi energy, temperature, and relaxation time, respectively,
$\overline{\omega}=1-i\omega\tau$, and $\xi=2\vert E_F\vert\tau/\hbar$.

The nonlinear optical response is described by nonlinear surface current densities determined by
second- and third-order nonlinear surface susceptibilities
\cite{cvs14newjphy,csa16acsnano,msa15njp,dh09apl,dh10prb,gla11jlet,mik11prb,ard14prb}. In the case
of THG, the third-order surface current density of graphene is expressed as:
\begin{equation} \label{eq:jTHG}
\mathbf{J}^{(3)}(\Omega_{3},\omega) = \bm{\sigma}_{s}^{(3)}(\Omega_{3};\omega)\vdots
\mathbf{E}(\omega)\mathbf{E}(\omega)\mathbf{E}(\omega),
\end{equation}
where $\Omega_{3}=3\omega$ is the frequency at the TH and $\bm{\sigma}_{s}^{(3)}$ is the
third-order nonlinear surface optical susceptibility. It is described by a single scalar function,
$\sigma_{s}^{(3)}$, via the relation
$\sigma_{s,ijkl}^{(3)}=\sigma_{s}^{(3)}(\delta_{ij}\delta_{kl}+\delta_{ik}\delta_{jl}+\delta_{il}\delta_{jk})/3$
\cite{cvs14newjphy,csa16acsnano}, with $\delta_{ij}$ being the Kronecker delta. Furthermore, in the
case of SHG arising from a local nonlinear polarization, the second-order nonlinear surface current
density can be written as:
\begin{equation}
\mathbf{J}^{(2)}(\Omega_{2},\omega) = \bm{\sigma}_{s}^{(2)}(\Omega_{2};\omega):
\mathbf{E}(\omega)\mathbf{E}(\omega),
\end{equation}
where $\Omega_{2}=2\omega$ is the frequency at the SH and $\bm{\sigma}_{s}^{(2)}$ is the
second-order nonlinear surface optical susceptibility. Symmetry considerations based on the fact
that graphene belongs to the $\mathcal{D}_{\mathrm{6h}}$ symmetry group lead to the conclusion that
the tensor $\bm{\sigma}_{s}^{(2)}(\Omega_{2};\omega)$ has three independent nonzero components,
$\sigma_{s,\perp\perp\perp}^{(2)}$,
$\sigma_{s,\parallel\parallel\perp}^{(2)}=\sigma_{s,\parallel\perp\parallel}^{(2)}$, and
$\sigma_{s,\perp\parallel\parallel}^{(2)}$, where the symbols $\perp$ and $\parallel$ refer to the
directions perpendicular onto and parallel to the plane of graphene, respectively. The values of
these parameters used in this study are
$\sigma_{s,\perp\perp\perp}^{(2)}=\SI[output-complex-root=\text{\ensuremath{i}}]{-9.71ie-16}{\ampere\meter\per\square\volt}$,
$\sigma_{s,\parallel\parallel\perp}^{(2)}=\sigma_{s,\parallel\perp\parallel}^{(2)}=\SI[output-complex-root=\text{\ensuremath{i}}]{-2.56ie-16}{\ampere\meter\per\square\volt}$,
and
$\sigma_{s,\perp\parallel\parallel}^{(2)}=\SI[output-complex-root=\text{\ensuremath{i}}]{-2.09ie-16}{\ampere\meter\per\square\volt}$
\cite{dh10prb,ard14prb}. Note that, as demonstrated in the SM, the qualitative conclusions of our
study do not change if instead of a local second-order nonlinear response of graphene one considers
a nonlocal one.
\begin{figure} [b!]\centering
\includegraphics[width=\columnwidth]{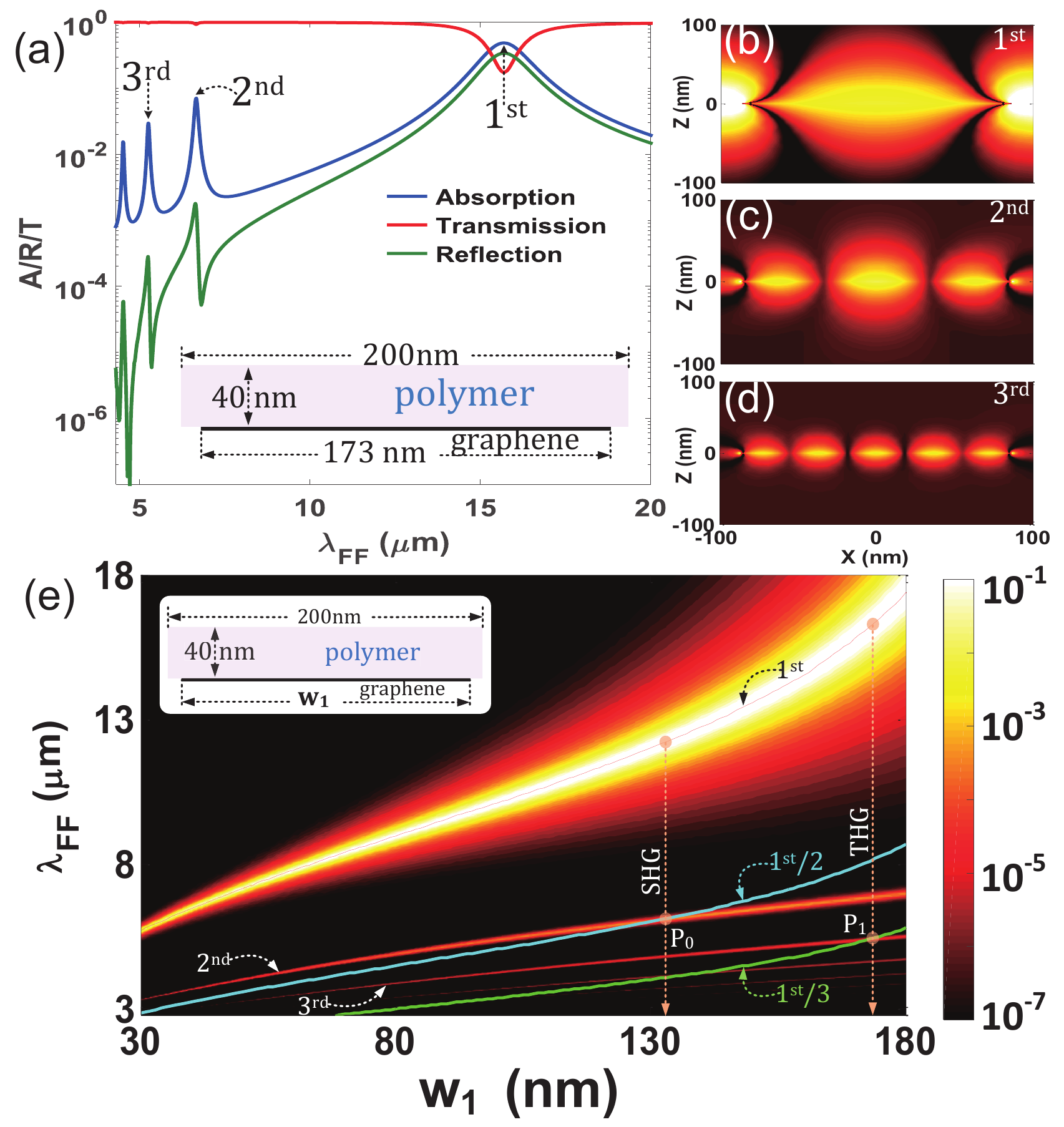}
\caption{(a) Wavelength dependence of the absorption, $A$, reflectance, $R$, and transmittance,
$T$. (b), (c), (d) Spatial profile of the dominant component of the electric field, $|E_x|$, at the
FF, determined for the first three SPP resonances, respectively. (e) Dispersion map of absorption
spectra vs. width of bottom GNRs. Yellow, blue, and green lines correspond to $\lambda_{FF}^{(1)}$,
$\lambda_{FF}^{(1)}/2$, and $\lambda_{FF}^{(1)}/3$, respectively, where $\lambda_{FF}^{(1)}$ is the
width-dependent wavelength of the first-order SPP.}\label{fig:OptimiBott}
\end{figure}

To characterize the linear optical response of the GIG structure, we first calculated the
absorption, $A$, transmittance, $T$, and reflectance, $R$, corresponding to the bottom graphene
grating with geometrical parameters given in the inset of \figref{fig:OptimiBott}(a), and with
$E_{F}=\SI{0.4}{\electronvolt}$, $\tau=\SI{0.2}{\pico\second}$, and $T=\SI{300}{\kelvin}$. The
results of these calculations are summarized in \figref{fig:OptimiBott}(a). It can be seen that the
absorption spectrum possesses a series of resonances, which are due to the excitation of SPPs on
the GNRs. The field distributions of the first three SPPs are given in
Figs.~\ref{fig:OptimiBott}(b)-\ref{fig:OptimiBott}(d), respectively. They show that the local
optical field is strongly enhanced and confined around GNRs, with the largest field enhancement
observed for the first-order SPP. Moreover, the results presented in \figref{fig:OptimiBott}(a)
show that the absorption and reflectance spectra have resonances at the same wavelengths, a feature
that is particularly useful for the optimization of the GIG structure.

A convenient procedure for designing a graphene grating in which a double-SPP-resonance phenomenon
occurs is illustrated by the dispersion map of the absorption at the FF, presented in
\figref{fig:OptimiBott}(e). The bands in this map, which show the width-dependent resonance
wavelengths of SPPs of different order, suggest that it is possible to choose the width $w_{1}$ in
such a way that a pair of SPPs exist at the FF and HH. Thus, if $w_{1}=\SI{132}{\nano\meter}$, a
double resonance exists at the FF and SH, i.e. at ($\lambda_{FF},\lambda_{SH}=\lambda_{FF}/2$),
with $\lambda_{SH}=\lambda(P_{0})=\SI{6.04}{\micro\meter}$, whereas if
$w_{1}=\SI{173}{\nano\meter}$, a double resonance exists at the FF and TH, i.e. at
($\lambda_{FF},\lambda_{TH}=\lambda_{FF}/3$), with
$\lambda_{TH}=\lambda(P_{1})=\SI{5.25}{\micro\meter}$.
\begin{figure}[t!]\centering
\includegraphics[width=\columnwidth]{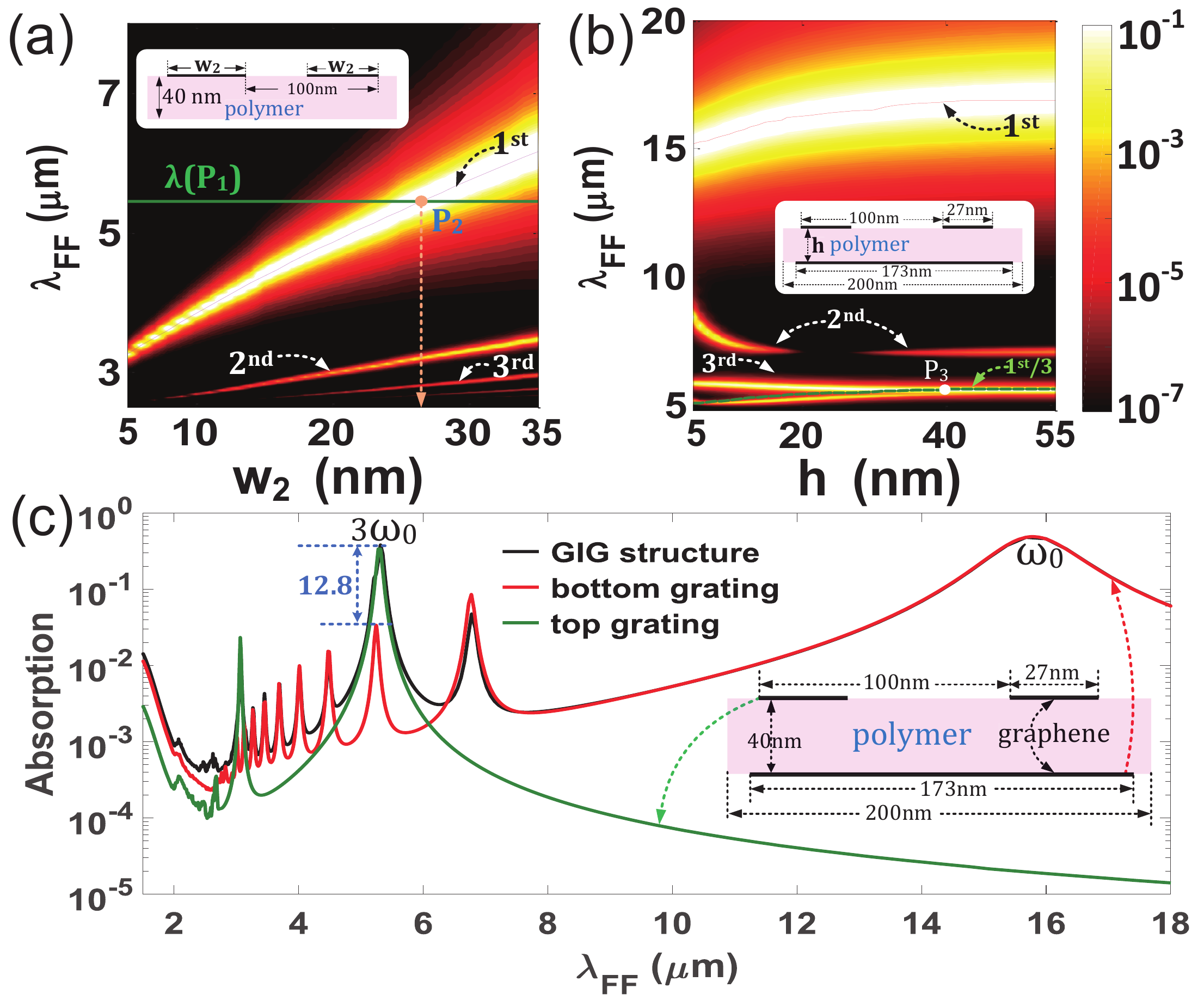}
\caption{(a) Dispersion map of the top graphene grating. The magenta line shows the width-dependent
wavelength of the first-order SPP. (b) Dependence of absorption spectra of the GIG structure on
$h$. Red and green lines correspond to $\lambda_{FF}^{(1)}$ and $\lambda_{FF}^{(1)}/3$,
respectively, where $\lambda_{FF}^{(1)}$ is the thickness-dependent wavelength of the first-order
SPP. (c) Absorption spectra of the optimized top and bottom gratings as well as that of the GIG
structure, determined for the optimal thickness $h=\SI{40}{\nano\meter}$ for which the GIG
structure possesses a double-resonance at frequencies $\omega_0$
($\lambda=\SI{15.9}{\micro\meter}$) and $3\omega_0$
($\lambda=\SI{5.3}{\micro\meter}$).}\label{fig:OptimiGIG}
\end{figure}

A drawback of the scheme we just described is that the plasmon at the HH is a \textit{higher-order
plasmon} and therefore it is less efficiently excited. In order to overcome this limitation and
further enhance the nonlinear optical response of the device, another graphene grating is placed
onto the spacer. The width $w_{2}$ of the GNRs of this top grating can be freely chosen. As such,
it is chosen in such a way that at the HH (SH or TH) \textit{first-order plasmons} exist in these
GNRs. For example, as illustrated in \figref{fig:OptimiGIG}(a), when $w_2=\SI{27}{\nano\meter}$ the
wavelength of the \textit{first-order plasmon} of the GNRs of the top grating is equal to
$\lambda(P_1)$. Therefore, we expect that when $w_{1}=\SI{173}{\nano\meter}$,
$w_2=\SI{27}{\nano\meter}$, and $h=\SI{40}{\nano\meter}$ the GIG structure possesses
\textit{first-order plasmons} at both the FF and TH. This property is verified by the dispersion
map of the absorption in the GIG structure, plotted in \figref{fig:OptimiGIG}(b). This map shows
that indeed the GIG structure has first-order plasmons at $\lambda_{FF}=\SI{15.9}{\micro\meter}$
and $\lambda_{TH}=\SI{5.3}{\micro\meter}$, predominantly localized at the bottom and top gratings,
respectively. Note that due to the optical coupling between the top and bottom gratings, the double
resonance phenomenon in the decoupled bottom grating appears at a pair of wavelengths slightly
blue-shifted as compared to those in the optimized GIG structure.

The optical coupling between the two gratings leads to several additional interesting phenomena, as
per \figref{fig:OptimiGIG}(b). First, the resonance wavelengths of SPPs vary with the thickness
$h$, especially at small values of $h$ for which there is a stronger coupling. Second, for
$h\lesssim\SI{40}{\nano\meter}$, the resonance wavelengths of the first-order plasmon of the top
grating and the third-order plasmon of the bottom grating are no longer equal, so that one expects
a smaller enhancement of the THG. On the other hand, if $h$ is too large, the electric field at the
FF in the bottom grating can no longer excite the first-order plasmon at the TH in the top grating,
which also leads to decreased enhancement of the THG. Therefore, the optimum value of $h$ is
$\sim$\SI{40}{\nano\meter}. Note also that for $\SI{21}{\nano\meter}<h<\SI{27}{\nano\meter}$, the
second-order plasmon in the GIG structure is almost completely suppressed, a phenomenon explained
by the fact that the system has a bound-state in the continuum for $h\simeq\SI{24}{\nano\meter}$
\cite{mbs08prl}.
\begin{figure}[b!]\centering
\includegraphics[width=\columnwidth]{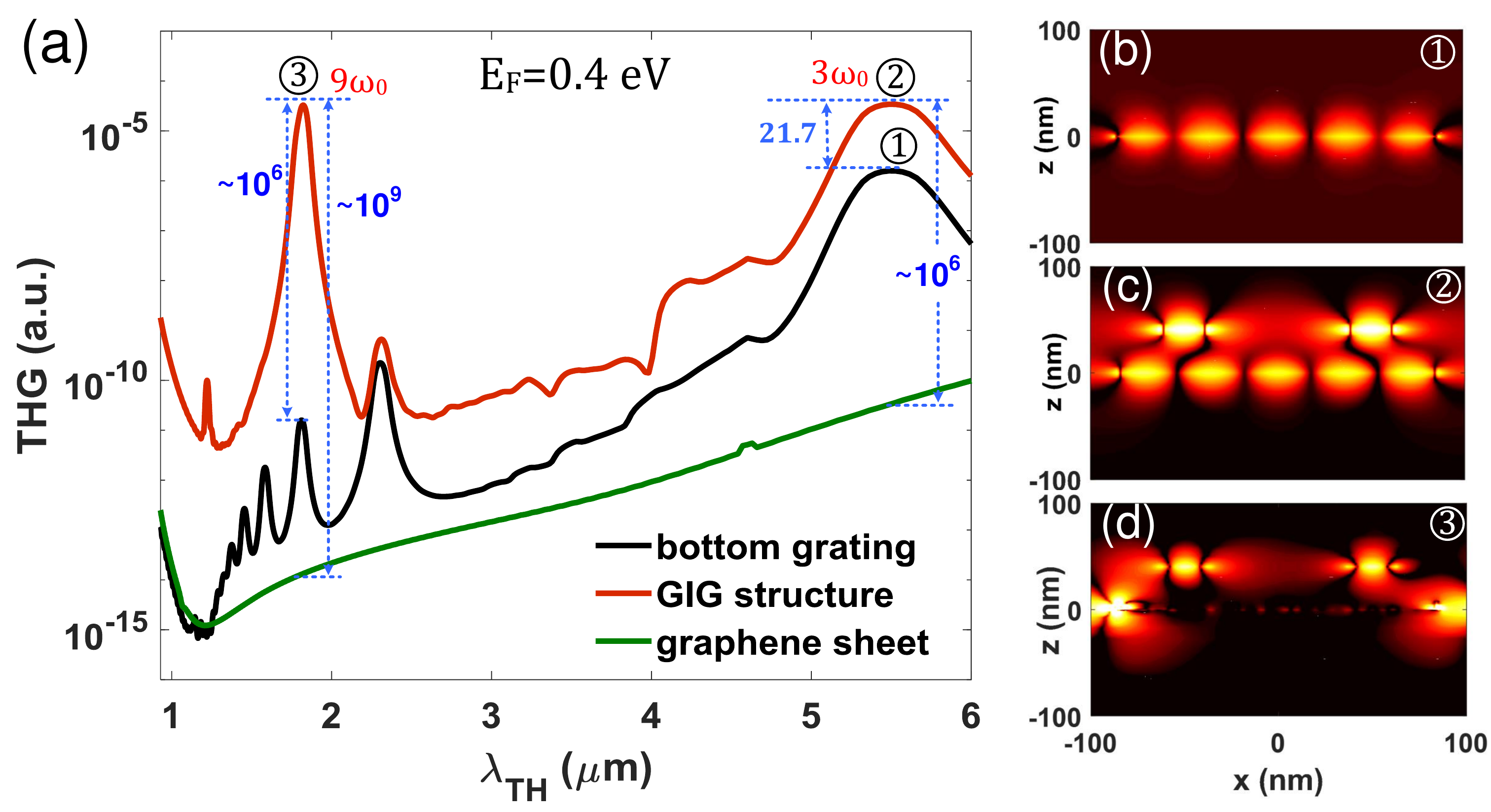}
\caption{(a) Spectra of THG for a graphene sheet, optimized bottom grating, and optimized GIG
structure. (b), (c), (d) Spatial profile of the dominant component of the electric field, $|E_x|$,
at the TH, determined for the resonances marked by \textcircled{1}, \textcircled{2}, and
\textcircled{3} in panel (a), respectively.}\label{fig:EnhanceTHG}
\end{figure}

These conclusions are further validated by the absorption spectra presented in
\figref{fig:OptimiGIG}(c), where we compare the absorption in the bottom grating optimized to
possess a double resonance at $\lambda_{FF}=\SI{15.75}{\micro\meter}$ and
$\lambda_{TH}=\lambda_{FF}/3=\SI{5.25}{\micro\meter}$, the absorption in the top grating designed
to possess a fundamental plasmon at the TH wavelength, $\lambda_{TH}=\SI{5.25}{\micro\meter}$, and
the absorption in the optimized GIG structure. These spectra show that by adding the top grating
the absorption at the TH is enhanced by more than 12 times, which suggests that the local optical
field and implicitly the nonlinear optical response of the GIG structure can be significantly
enhanced.

To quantify the enhancement of the THG in our GIG structure, we computed the THG spectra for a
graphene sheet, the bottom grating optimized to possess a double resonance at
$\lambda_{FF}=\SI{15.75}{\micro\meter}$ and $\lambda_{TH}=\lambda_{FF}/3=\SI{5.25}{\micro\meter}$,
and the optimized GIG structure, the results being compared in \figref{fig:EnhanceTHG}(a). These
spectra show that, as compared to the graphene sheet, the THG in the optimized bottom grating is
enhanced by $\sim$$10^{5}$ when the FF coincides with that of the first-order plasmon of the bottom
GNRs. Under the same excitation conditions, an additional $21\times$ enhancement is observed in the
GIG structure. These results are explained by the spatial profiles of the amplitude of the dominant
component of the TH electric field, $E_{x}$, presented in Figs.~\ref{fig:EnhanceTHG}(b) and
\ref{fig:EnhanceTHG}(c). Thus, in the optimized bottom grating, at the TH, a third-order plasmon is
excited, whereas in the optimized GIG structure both a first-order plasmon of the top grating and a
third-order plasmon of the bottom grating are generated. Importantly, it can be seen that when the
FF [$3\omega_{0}$ in \figref{fig:EnhanceTHG}(a)] is equal to that of the first-order plasmon of the
top grating and the third-order plasmon of the bottom grating the THG is enhanced by
$\sim$$10^{9}$, as compared to the case of a graphene sheet.
\begin{figure}[b!]\centering
\includegraphics[width=\columnwidth]{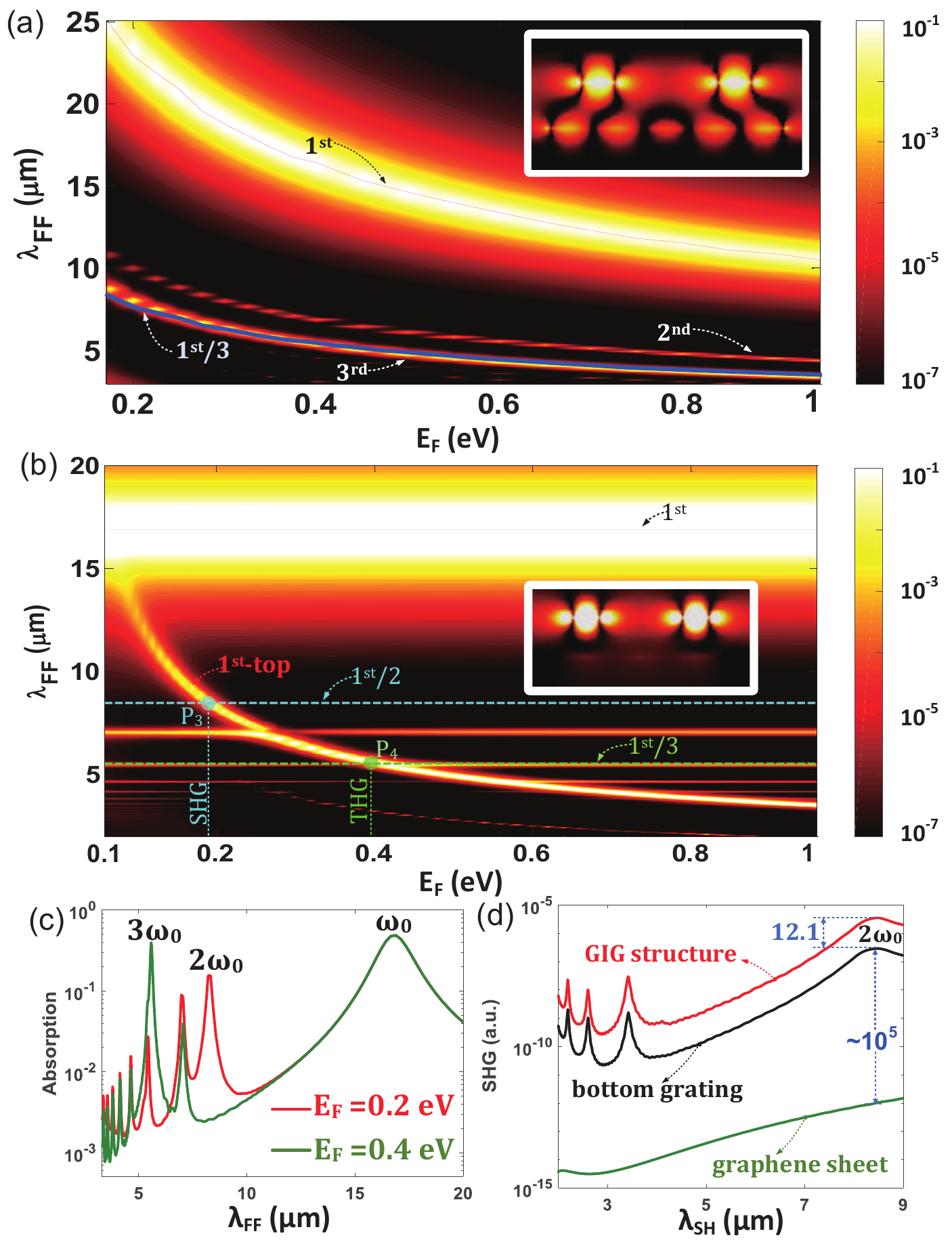}
\caption{(a) Absorption spectra of the optimized GIG structure vs. the Fermi energy of the two
graphene gratings. Red and blue lines correspond to $\lambda_{FF}^{(1)}$ and
$\lambda_{FF}^{(1)}/3$, respectively, where $\lambda_{FF}^{(1)}$ is the Fermi-energy-dependent
wavelength of the first-order SPP. In inset, profile of the TH electric field, $|E_x|$, determined
for $E_{F}=\SI{0.3}{\electronvolt}$. (b) The same as in (a), but calculated for the case when only
$E_{F}$ in the top grating varies and $E_{F}=\SI{0.4}{\electronvolt}$ in the bottom grating. In
inset, profile of the SH electric field, $|E_x|$, determined for $E_{F}=\SI{0.3}{\electronvolt}$.
(c) Absorption spectra of a GIG structure optimized to enhance SHG
($E_{F}=\SI{0.2}{\electronvolt}$) and THG ($E_{F}=\SI{0.4}{\electronvolt}$). (d) Spectra of SHG
determined for a graphene sheet placed on a polymer substrate, the bottom grating, and the
optimized GIG structure.}\label{fig:EnhanceSHG}
\end{figure}

A particularly important property of the proposed GIG structure is the broadband nonlinearity
enhancement at the HH, achievable by tuning the Fermi energy in the two gratings. The reason for
this unique property is revealed by the dispersion map of the absorption of the optimized GIG
structure, presented in \figref{fig:EnhanceSHG}(a). Thus, it is clear from this figure that the
ratio between the wavelengths of the first-order SPP of bottom GNRs on the one hand, and
third-order SPPs of the bottom GNRs and first-order SPPs of the top GNRs on the other hand, remains
constant as the Fermi energy varies. Consequently, the double-resonance property is precisely
preserved as the Fermi energy varies. More specifically, as shown in \figref{fig:EnhanceSHG}(a),
when the Fermi energy is varied from \SIrange{0.2}{1}{\electronvolt}, the resonance wavelength of
the first-order plasmon of bottom GNRs, and implicitly the operating wavelength at the FF, varies
from \SIrange{25}{10}{\micro\meter}.

Another remarkable property of our proposed GIG structure is that it can enhance both the THG and
SHG. Specifically, this functionality can be realized by tuning the Fermi energy only in the top
grating, such that the resonance wavelength of first-order SPPs of the GNRs in this grating is
shifted from the TH to the SH. This is demonstrated by the absorption map of the GIG structure
presented in \figref{fig:EnhanceSHG}(b). Thus, this figure shows two types of plasmon bands, namely
flat bands corresponding to SPPs in the bottom grating, which obviously do not depend on $E_{F}$ in
the top grating, and plasmon bands associated to the top grating, whose resonance wavelength
depends on $E_F$. In particular, it can be seen that whereas the resonance wavelength on the
first-order SPPs of the bottom grating remains constant, $\lambda_{FF}=\SI{15.75}{\micro\meter}$,
the resonance wavelength of first-order SPPs of the top grating varies from
$\lambda(P_{4})=\lambda_{FF}/3=\SI{5.25}{\micro\meter}$ to
$\lambda(P_{3})=\lambda_{FF}/2=\SI{7.875}{\micro\meter}$ when $E_{F}$ is tuned from
\SIrange{0.4}{0.2}{\electronvolt}, respectively [see also \figref{fig:EnhanceSHG}(c)].

The strong enhancement of the SHG of the GIG structure, achieved for
$E_{F}=\SI{0.2}{\electronvolt}$, is clearly demonstrated by the plots presented in
\figref{fig:EnhanceSHG}(d), where we show the SHG spectra corresponding to a graphene sheet placed
on the polymer substrate, the bottom grating, and the combined GIG structure, determined for
$E_{F}$ for which the top GNRs have first-order SPPs at the SH. As in the case of the TH, one can
see that strongly enhanced SHG can be achieved in the optimized GIG structure. In particular, at
resonance, the SHG in the bottom grating is $\sim$$10^{5}$ larger than in the case of a graphene
sheet, whereas a further order of magnitude enhancement is achieved in the optimized GIG structure.

To conclude, a highly engineered GIG metasurface for enhancement of SH and TH is studied in this
Letter. We demonstrate that it can be used to achieve tunable and dual-broadband enhancement of
both nonlinear optical interactions, a property originating from the fact that our system possesses
tunable double-resonances. In practice, this nonlinearity enhancement can be further improved by
stacking several GIG units together to construct a 3D graphene metamaterial \cite{ylc12natnano}.
This new type of graphene structures could open up new research directions towards the development
of novel ultra-compact and multi-frequency active photonic nanodevices.

\begin{acknowledgments}
The authors acknowledge the use of UCL Legion High Performance Computing Facility (Legion@UCL), and
associated support services, in the completion of this work. This work was supported by European
Research Council (ERC), Grant Agreement no. ERC-2014-CoG-648328. \end{acknowledgments}

\end{document}